\documentclass{article}         
\usepackage{graphicx}
\usepackage[figuresright]{rotating}
\usepackage{amsmath} 
\usepackage{amssymb} 
\usepackage{amsfonts} 
\usepackage{rotating}
\usepackage{subfigure}

\begin{document}



\title{Time-varying clustering of multivariate longitudinal observations}

\author{Antonello Maruotti\footnote{Southampton Statistical Sciences Research Institute \& School of Mathematics, University of Southampton, Building 39, Southampton SO17 1BJ UK, email: a.maruotti@soton.ac.uk}\footnote{Dipartimento di Scienze Politiche, Universit\`{a} di Roma Tre, Via G. Chiabrera, 199 - 00145 Roma, Italy, email: antonello.maruotti@uniroma3.it}
\and Maurizio Vichi\footnote{Dipartimento di Scienze Statistiche, Sapienza Universit\`{a} di Roma, Roma, Italy, email: maurizio.vichi@uniroma1.it}}

\date{}

\maketitle

\begin{abstract}
We propose a statistical method for clustering of multivariate longitudinal data into homogeneous groups. This method relies on a time-varying extension on the classical K-means algorithm, where a multivariate vector autoregressive model is additionally assumed for modeling  the evolution of clusters' centroids over time. We base the inference on a least squares specification of the model and coordinate descent algorithm. To illustrate our work, we consider a longitudinal dataset on human development. Three variables are modeled, namely life expectancy, education and gross domestic product.
\end{abstract}

%


\maketitle


%

\section{Introduction}
\label{s:intro}
This paper considers estimation and inference in multivariate models for longitudinal data within a time-varying clustering framework. Longitudinal data typically refer to data containing time-series observations of a number of statistical units. Therefore, observations in longitudinal data involve at least two dimensions: a cross-sectional dimension and a time-series dimension. There are at least three factors contributing to the growth of longitudinal data studies: data availability (see, e.g. Wooldridge, 2002; Fitzmaurice et al., 2008; Baltagi, 2009); greater capacity of modelling the complexity of phenomena under study than a single cross-section or time-series data; challenging methods. Indeed, longitudinal data, by blending the inter-unit differences and intra-unit dynamics, have several advantages over cross-sectional or time-series data. Containing more variability than other types of data, longitudinal data improves the efficiency of statistical estimates, by providing more accurate inference of model parameters. Moreover, more complicated behavioral hypothesis can be constructed and tested, as well as the impact of latent variables can be controlled, uncovering hidden dynamic relationships. 

New challenges appear when analyzing longitudinal data within a clustering framework. Clustering methods generally aim at partitioning units into meaningful clusters, measuring the homogeneity within clusters as well as the difference between clusters. Standard clustering approaches have been considerably improved, allowing for solutions to some practical issues such as the choice of the number of clusters, the allocation to clusters and the clustering algorithm obtained. Several approaches have been developed in model-based and non-model-based frameworks. In a model based framework, related to mixture modelling techniques, examples aiming at clustering longitudinal data are of various nature (see e.g. De la Cruz-Mesia et al., 2008; McNicholas and Murphy, 2010; Alf\`{o} and Maruotti, 2010, Maruotti, 2011; Komarek and Komarkova, 2013; Ciampi et al., 2013). Fewer methodologies have been developed in a non-model-based framework in the literature (D'Urso, 2004; Tarpey, 2007; Genolini and Falissard, 2010).

We would contribute to extend this branch of literature by introducing a new method to cluster multivariate longitudinal observations focusing on the evolution of partitions over time and taking into account latent heterogeneity and time-dependence, simultaneously. In an univariate longitudinal setting, an attempt at describing processes dynamics can be achieved by modeling the current value of the outcome as a weighted linear sum of its previous values. This is an autoregressive process and is a very simple, yet effective, approach to account for time-dependence structure. 
However, in many applications the univariate setting may be restrictive since only partially captures the inter-relationships among different aspects of the analyzed phenomenon and, furthermore, it is less efficient than the multivariate approach whenever outcomes show a certain degree of correlation. Thus, a multivariate model should be adopted. 
To account for the components that need to be described by a model for multivariate longitudinal data, i.e. serial dependence that arise due to the nature of repeated measurements over time and heterogeneity in the observed units, we propose a multivariate vector autoregressive (MVAR) model in a clustering framework. MVAR is a flexible and easy to use model for the analysis of dynamic multivariate time-series. We extend MVAR under a longitudinal setting where multivariate time-series are observed for each unit of the longitudinal study with the aim of describing the dynamic behavior and clustering the different time-series (other approaches in a time series framework can be found in Coppi et al., 2010; Maharaj and D'Urso, 2011). The resulting model allows for heterogeneity and \textit{dynamic clustering}, modeled by a MVAR on the units' partitions centroids. The proposed approach has some potential advantages over competitive methods. It does not require any distributional assumption within clusters. It is an interesting by-product when no prior information on the clustering structure is available. A simple structured assumption regarding the shape of the clustering dynamics is made. Robustness with respect to the computational burden involved in the estimation step and invariance to time-scaling are other properties of the proposed approach. Furthermore, it is possible to evaluate the trajectory of each cluster and cluster memberships are time-varying. The model parameters are estimated via least squares (LS) by using a coordinate descent algorithm. 

The proposal is illustrated on real data by analyzing a sample taken for the United  Nations Development Program (UNDP) database where life expectancy, education and GDP indexes used  in the computation of the human development index (HDI) are considered as response variables. Data are recorded for 143 Countries over 9 years (from 1997 to 2005). The three dimensions in the HDI represent the minimum set of 
indicators for assessing the aggregate level of human development. The HDI combines indicators of very different sensitivity to change and as such it can hardly capture the short-term improvements: social indicators of longevity and literacy (stock), strictly 
related to demographics; the slow social indicator on gross education enrollment; the fast changing economic flow indicator on living standard. These three dimensions address conceptually different aspects of human development, which although correlated do not 
predetermine one another. We provide evidence of the usefulness of a multivariate approach and of the presence of a latent time-varying structure. A clustering structure is identified corresponding to different levels of development, as well as a dynamic process describing the evolution of human development. 

{The paper is organized as follows. In Section \ref{s:model} we introduce our approach for multivariate dynamic classification, by specifying the adopted notation, by providing least square estimation and a model selection criterion. In Section \ref{s:appl}, we describe the data and analyze HDI components in a univariate and multivariate setting by looking at clustering and time dependence. A final discussion follows in Section \ref{s:disc}.}

\section{Partitioning longitudinal multivariate observations}\label{s:model}

\subsection{Notation}\
For the convenience of the reader the notation and the terminology used is listed here.
\begin{description}
\item $n,J,T,G,P$: number of units, variables, times, classes of units; number of lags of order of the VAR model.
\item[{X}, {E}]: ($n\times J \times T$) three-way data and three-way array of error terms; where $x_{ijt}$ (similarly for $e_{ijt}$) is the value (error) of the $j$-th variable on the $i$-th unit at time $t$.
\item[U] = [$u_{igt}$]: ($n\times G\times T$), $[\textbf{U}_{..1},\textbf{U}_{..2},\dots,\textbf{U}_{..T}]$ binary matrix specifying a partition of the units $\textbf{U}_{..t}$ for each time $t$, where $u_{igt}=1$ if the $i$-th unit belongs to class $g$ at time $t$, $u_{igt}=0$ otherwise. Matrix $\textbf{U}$ has only one nonzero element per row.
\item[$\overline{\textbf{X}}$] = [$\bar{x}_{gjt}$]: ($G\times J\times T$), 
[$\bar{\textbf{X}}_{..1}$,$\bar{\textbf{X}}_{..2}$,\dots,$\bar{\textbf{X}}_{..T}$] matrix of centroids $\bar{\textbf{X}}_{..t}$ for each time $t$ including the $G$ class-conditional mean vectors $\bar{\textbf{x}}_{g.t}$ in the $J$-space.
\end{description}

\subsection{The model}

The partition of the units over $T$ times can be achieved by estimating the parameter matrices $\textbf{U}_{..t}$ and $\bar{\textbf{X}}_{..t}$ of the following model

\begin{equation}
\textbf{x}_{i.t} = \bar{\textbf{X}}_{..t}'\textbf{u}_{i.t}+\textbf{e}_{i.t}
\end{equation}
and rewritten in compact form
\begin{equation}\label{mod}
\textbf{X}_{..t} = \textbf{U}_{..t}\bar{\textbf{X}}_{..t}+\textbf{E}_{..t}, \quad t = 1,\dots, T.
\end{equation}

The dynamic evolution of a partition of units along time for each time can be modeled by considering the following autoregressive VAR(P) model on the centroids of the partitions
\begin{equation}
\boldsymbol{\bar{x}}_{g.t} = \boldsymbol{c} + \boldsymbol{A}_1\boldsymbol{\bar{x}}_{g.t-1}+ \boldsymbol{A}_2\boldsymbol{\bar{x}}_{g.t-2}+\dots+ \boldsymbol{A}_P\boldsymbol{\bar{x}}_{g.t-P}+\boldsymbol{w}_{g.t}
\end{equation}
where \textbf{c} denotes a ($J\times 1$) vector of constants, $\textbf{A}_p$ is a ($J\times J$) matrix of autoregressive coefficients for $p=1,\dots,P$. The ($J\times 1$) vector $\textbf{w}_{g.t}$ is a white noise process such that $E[\textbf{w}_{g.t}] = 0$, and
\begin{equation*}
E[\textbf{w}_{g.t},\textbf{w}_{g.t'}] = \left\{\begin{array}{cc}\boldsymbol{\Omega},& t = t'\\0,& {\rm otherwise}\end{array}\right. .
\end{equation*}
 Rewritten in compact form we have
\begin{equation}
\boldsymbol{\bar{X}}_{..t} = \textbf{1}_G\textbf{c}' + \boldsymbol{\bar{X}}_{..t-1}\textbf{A}_1'+ \boldsymbol{\bar{X}}_{..t-2}\textbf{A}_2'+\dots+ \boldsymbol{\bar{X}}_{..t-P}\textbf{A}_P'+\textbf{W}_{..t}
\end{equation}
where $\textbf{W}_{..t} = [\textbf{w}_{g.p},\textbf{w}_{g.p+1},\dots,\textbf{w}_{g.T}]$

Now we are in position to specify the K-clustering autoregressive VAR(P) model for modeling the dynamic evolution of the clusters. We include an AR(P) model as specified in equation (4) into the clustering model (2). Formally,

\begin{eqnarray}\label{arp}
\textbf{X}_{..t} &=& \textbf{1}_N\textbf{c}'+\textbf{U}_{..t}\bar{\textbf{X}}_{..t-1}\textbf{A}_1'+\textbf{U}_{..t}\bar{\textbf{X}}_{..t-2}\textbf{A}_2'+\dots+
\textbf{U}_{..t}\bar{\textbf{X}}_{..t-p}\textbf{A}_p'+\textbf{D}_{..t},
\end{eqnarray} 

where $$\textbf{D}_{..t} = \textbf{U}_{..t}\textbf{W}_{..t}+\textbf{E}_{..t}$$

This model will be named CAR(K,P), i.e. K-clustering P-autoregressive model.
\subsection{Least squares estimation}
Let us start to consider the CAR(K,1) model. To evaluate the partition of the objects alongside, the AR(1) is included in the partitioning model (\ref{mod}). Thus, we have 
\begin{eqnarray}\label{ar1}
\textbf{X}_{..t} &=& \textbf{1}_N\textbf{c}'+\textbf{U}_{..t}\bar{\textbf{X}}_{..t-1}\textbf{A}_1'+\textbf{D}_{..t},\quad t = 2,\dots,T
\end{eqnarray} 
or in more compact form
\begin{equation}
\textbf{X}_{..t} = \textbf{U}_{..t}\dot{\bar{\textbf{X}}}_{..t-1}\textbf{B}_1'+\textbf{D}_{..t},\quad t = 2,\dots,T
\end{equation}
where $\textbf{B}_1 = [\textbf{c},\textbf{A}_1]$ and $\dot{\bar{\textbf{X}}}_{..t-1}= [\textbf{1}_G, \bar{\textbf{X}}_{..t-1}]$.

The least-squares estimation of the model (\ref{ar1}) can be obtained  by solving
\begin{eqnarray}\label{ols}
F(\textbf{B}_1,\textbf{U}_{..t},\bar{\textbf{X}}_{..t-1}) &=& \sum_{t=2}^T ||\textbf{X}_{..t}-\textbf{U}_{..t}\dot{\bar{\textbf{X}}}_{..t-1}\textbf{B}_1'||^2\nonumber\\ &=& \sum_{t=2}^T||\textbf{X}_{..t}-\textbf{1}_N\textbf{c}'- \textbf{U}_{..t}\dot{\bar{\textbf{X}}}_{..t-1}\textbf{A}_1'||^2\rightarrow \operatorname*{\min_{\textbf{U}_{..t}\dot{\bar{\textbf{X}}}_{..t-1}\textbf{A}_1}}
\end{eqnarray}
subject to $\textbf{U}$ being a binary and row-stochastic matrix.

Problem (\ref{ols}) can be solved by considering a coordinate descent algorithm (see Zangwill, 1969), which updates parameters $\textbf{c},\textbf{U}_{..t}\dot{\bar{\textbf{X}}}_{..t-1}\textbf{A}_1$ and decreases the loss function. Each matrix to be estimated is updated while maintaining all the others fixed. 

Thus, given $\hat{\textbf{U}}_{..t}$ and $\hat{\textbf{B}}_{1}$, the optimal $\bar{\textbf{X}}_{..t}$ is specified solving a multivariate regression problem
\begin{equation}
\hat{\dot{\bar{\textbf{X}}}}_{..t-1} = (\hat{\textbf{U}}_{..t}'\hat{\textbf{U}}_{..t})^{-1}\hat{\textbf{U}}_{..t}'\textbf{X}_{..t}\hat{\textbf{B}}_{1}(\hat{\textbf{B}}_{1}'\hat{\textbf{B}}_{1})^{-1}.
\end{equation}

Given $\hat{\dot{\bar{\textbf{X}}}}_{..t-1}$ and $\hat{\textbf{U}}_{..t}$, for the optimal $\textbf{B}_1$ the loss function (\ref{ols}) can be written as 

\begin{equation*}
F(\textbf{B}_1) = \sum_{t=2}^T||\textbf{X}_{..t}-\hat{\textbf{U}}_{..t}\hat{\dot{\bar{\textbf{X}}}}_{..t-1}\textbf{B}_1'||^2 = \left\Arrowvert\left(\begin{array}{c}\textbf{X}_{..2}\\ \vdots \\ \textbf{X}_{..T}\end{array}\right)-\left( \begin{array}{c}\hat{\textbf{U}}_{..1}\hat{\dot{\bar{\textbf{X}}}}_{..t-1}\\ \vdots \\ \hat{\textbf{U}}_{..T}\hat{\dot{\bar{\textbf{X}}}}_{..T-1}\end{array}\right)\textbf{B}_1'\right\Arrowvert
\end{equation*}
which is a straightforward application of a multivariate regression problem with solution

\begin{equation}\textbf{B}_1' = \left(\sum_{t=2}^T\hat{\dot{\bar{\textbf{X}}}}_{..t-1}'\hat{\textbf{U}}_{..t}'\hat{\textbf{U}}_{..t}\hat{\dot{\bar{\textbf{X}}}}_{..t-1}\right)^{-1}\sum_{t=2}^T\dot{\bar{\textbf{X}}}_{..t-1}'\hat{\textbf{U}}_{..t}\textbf{X}_{..t}
\end{equation}

Given $\hat{\dot{\bar{\textbf{X}}}}_{..t-1}$ and $\hat{\textbf{B}}_1$, the optimal $\textbf{U}_{..t}$ is given by solving an assignment problem

\begin{equation}\left\{\begin{array}{cc}
u_{igt} = 1, & {\rm if \quad}||\textbf{x}_{i.t}-\hat{\textbf{B}}_1\hat{\dot{\bar{\textbf{x}}}}_{g.t-1}||^2 = \min\{||\textbf{x}_{i.t}-\hat{\textbf{B}}_1\hat{\dot{\bar{\textbf{x}}}}_{l.t-1}||^2 : l=1,\dots,G\}\\
u_{igt} = 0, & {\rm otherwise}
\end{array}\right.\end{equation}

For the general CAR(K,P) least square estimation, we can follow similar steps as for CAR(K,1) model.
Let us rewrite the CAR(K,P) model (5) in compact form.

\begin{equation}
\textbf{X}_{..t} = \textbf{U}_{..t}\dot{\bar{\textbf{X}}}_{..t-P}\textbf{B}_P'+\textbf{D}_{..t},\quad t = 2,\dots,T
\end{equation}

where $\textbf{B}_P = [{\bf c},{\bf A}_1, {\bf A}_2,\dots,{\bf A}_P]$ and $\dot{\bar{\textbf{X}}}_{..t-P} = [{\bf 1}_G, \bar{\bf X}_{..t-1},\dots,\bar{\bf X}_{..t-P}].$

The least square estimation can be given minimizing the quadratic function
\begin{equation}
F(\textbf{B}_P,\textbf{U}_{..t},\dot{\bar{\textbf{X}}}_{..t-P}) = \sum_{t=2}^T||\textbf{X}_{..t}-\textbf{U}_{..t}\dot{\bar{\textbf{X}}}_{..t-P}\textbf{B}_P'||^2 
\end{equation}

with respect to $\textbf{B}_P,\textbf{U}_{..t},\dot{\bar{\textbf{X}}}_{..t-P}$ subject to $\textbf{U}_{..t}$ being binary and row stochastic. Thus, given the estimates $\hat{\textbf{U}}_{..t}$ and $\hat{\bf{B}}_P$ the optimal $\dot{\bar{\textbf{X}}}_{..t-P}$ is

\begin{equation}
\hat{\dot{\bar{\textbf{X}}}}_{..t-P} = (\hat{\textbf{U}}_{..t}'\hat{\textbf{U}}_{..t})^{-1}\hat{\textbf{U}}_{..t}'\textbf{X}_{..t}\hat{\textbf{B}}_{P}(\hat{\textbf{B}}_{P}'\hat{\textbf{B}}_{P})^{-1}.
\end{equation}

and each centroids matrix $\bar{\bf{X}}_{..t-l}$ is 

\begin{equation}
\bar{\bf{X}}_{t-l} =\left (\hat{\textbf{U}}_{..t}'\hat{\textbf{U}}_{..t}\right)^{-1}\hat{\textbf{U}}_{..t}'\left(\textbf{X}_{..t}-\textbf{1}_N{\bf c}'\sum_{p=1}^P\hat{\textbf{U}}_{..t}\bar{\bf{X}}_{t-p}{\bf A}_p'\right){\bf A}_l \left({\bf A}_l'{\bf A}_l\right)^{-1}.
\end{equation}

Given $\hat{\dot{\bar{\textbf{X}}}}_{..t-P}$ and $\hat{\textbf{U}}_{..t}$ the optimal $\bf{B}_P$ is

\begin{equation}
\hat{\bf{B}}_P = \left(\sum_{t=2}^T\hat{\dot{\bar{\textbf{X}}}}_{..t-P}\hat{\textbf{U}}_{..t}'\hat{\textbf{U}}_{..t}\hat{\dot{\bar{\textbf{X}}}}_{..t-P}\right)^{-1}\sum_{t=2}^T\hat{\dot{\bar{\textbf{X}}}}_{..t-P}\hat{\textbf{U}}_{..t}\bf{X}_{..t}.
\end{equation}

Finally, given $\hat{\dot{\bar{\textbf{X}}}}_{..t-P}$ and $\hat{\bf{B}}_P$, the optimal ${\bf U}_{..t}$ can be obtained by solving 

\begin{equation}\left\{\begin{array}{cc}
u_{igt} = 1, & {\rm if \quad}||\textbf{x}_{i.t}-\hat{\textbf{B}}_P\hat{\dot{\bar{\textbf{x}}}}_{g.t-P}||^2 = \min\{||\textbf{x}_{i.t}-\hat{\textbf{B}}_P\hat{\dot{\bar{\textbf{x}}}}_{l.t-P}||^2 : l=1,\dots,G\}\\
u_{igt} = 0, & {\rm otherwise}
\end{array}\right.\end{equation}

\subsection{Coordinate descent algorithm for CAR(K,P)}
The coordinate descent algorithm alternates three steps starting from a set of random or rational partitions ${\bf U}_{..t}, t = 1,\dots,T.$
\begin{description}
\item[Step 1:] Update centroids $\hat{\dot{\bar{\textbf{X}}}}_{..t-P}$ by (14).
\item[Step 2:] Update autoregressive coefficients ${\bf B}_P$ by (16).
\item[Step 3:] Update partitions ${\bf U}_{..t}$ by (17).
\end{description}

At each step, function (13) decreases or at least does not increase. Since (13) is bounded by below, the algorithm stops to a solution when (13) decreases less than an arbitrary fixed constant. The solution is at least a local minimum of the problem, thus, the researcher is advised to repeat the algorithm from different initial solutions retaining the best one in order to increase the chance to find the global minimum of the problem. To avoid local maxima, we run the algorithm 10 times, and retain the {\it best} solution on the basis of the selection criterion described in the next subsection.

\subsection{Model selection}
Without a measure for model selection, it may be difficult to correctly assess model behavior and quality. Here, the model selection is based on an internal criterion defined solely from the data that evaluates each partition of the observations into groups through some function measuring adequacy of each partition. Clustering criteria are often based on decomposition of the total variance of the data into the total within-group variance and the total between-group variance. A natural criterion for the evaluation of a certain partition is the amount of heterogeneity explained by the obtained partition. Thus, to chose the optimal number of clusters , we use the Calinski and Harabasz criterion $CH(G)$ (Calinski and Harabasz, 1974). The $CH(\cdot)$ criterion combines the within and the between covariance matrices to evaluate the clustering quality. The within covariance matrix is 

$$W = \sum_{g=1}^G\sum_{t=1}^T\sum_{i=1}^n({\bf x}_{i.tg}-\bar{{\bf x}}_{g.t})({\bf x}_{itg}-\bar{{\bf x}}_{g.t})'$$
and the between covariance is 

$$B = \sum_{g=1}^G\sum_{t=1}^Tn_{gt}(\bar{\bf{x}}_{g.t}-\bar{\bf{x}}_{..t})(\bar{\bf{x}}_{g.t}-\bar{\bf{x}}_{..t})'$$

where ${\bf x}_{i.tg}$ collects data clustered in $g$ at time $t$ and $n_{gt}$ represents the number of units in cluster $g$ at time $t$. The optimal number of clusters corresponds to the value of $G$ that maximizes

$$CH(G) = \frac{trace(B)}{trace(W)}\frac{n\times T-G}{G-1}$$

Its computation is very fast and, therefore, it is possible to run our method repeatedly with different starting points and to use the solution with the largest value of $CH(\cdot)$.


\section{Empirical application: the Human Development Index}\label{s:appl}

\subsection{Data description}
The empirical application is derived from the analysis of the HDI for 143 Countries over 9 years (from 1997 to 2005). In 1990 the United Nations Development Programme introduced a new way of measuring development by combining indicators into a composite development index, the HDI. The breakthrough for the HDI was the creation of a single index which was to serve as a frame of reference for both social and economic development. It is composed from statistics derived by life expectancy index (LEI), education index (EI) and gross domestic product index (GDPI) collected at the national level. The HDI represents the uniformly weighted sum of: the life expectancy (LE) at birth, measured as an index of population health and longevity; the education, measured by the adult literacy rate (ALR, with 2/3 weighting) with gross enrollment ratio (GER, with 1/3 weighting); the GDPI as a measure of standards of living given by the GDP per capita at purchasing power parity. Each of HDI components is mapped onto a unit free index. In detail, each HDI component is given by

$$Index_{ijt} = \frac{x_{ijt}-\max(x_{ijt})}{\max(x_{ijt})-\min(x_{ijt})}$$

HDI, although frequently used when synthesizing the degree of development of a Country, presents an evident limit since it is built as a simple arithmetic mean of three social-economic indicators. Therefore, a novel formulation of the HDI has been proposed in recent literature, characterized by a different definition of its three components and presented by their geometric mean. Nevertheless, in this application we consider the non-updated version of the index as an illustrative example. Furthermore, data refer to a period where the {\it old} version of the HDI was applied. The proposed methodology can be straightforwardly applied to the new version of the index.

However, studying and monitoring human development often involve the analysis of multivariate longitudinal data, i.e. a number of different variables measured over time on a set of Countries. The growing interest in human development has led to a substantial literature. One branch of this is concerned with synthetic indexes designed to summarize a series of measurements and to provides easily interpretable results for widespread dissemination (see e.g. Konya and Guisan, 2008; Nathan and Mishra, 2010). On the other hand, synthetic index as HDI may fail to take into account certain aspects of development processes that deserve particular attention. These include the association between indicators composing the HDI (for a discussion on this topic see e.g. Cahill, 2005) and the serial dependence structure in the longitudinal data, as well as the unobserved heterogeneity sources that typically arise in these studies. Due to these limitations (for a critical review of the HDI see e.g. Sagar and Najam, 1998) a different approach should be pursued. 

This dataset contains sufficient data to enable the determination of development patterns and changes in Countries behaviors over time. We observe Countries for all 5 continents. 
The range of each time-series varies considerably across Countries and over time. For example the overall median for LEI is 0.760 but can rise 0.954 in some Countries (e.g Japan) and can show very low values, such as in Swaziland with a value of 0.100. Similarly for EI and GDPI, we observe that in Western Countries both levels are quite high. For the EI e.g. Australia, New Zeland, Denmark, Finland and Ireland reach a value of 0.993, while Bhutan and Guinea show the lowest value at 0.330 with respect to a median value at 0.860. As expected, poor Countries (most of Africans) show very low values for the GDPI, e.g. for Rwanda we observe 0.310, while some economic developed Countries may reach a value of 1, e.g. Luxembourg, Norway, USA, Ireland, while the median level is 0.680. Descriptive statistics (see Table 1) provide few insights but not exhaustive results. A first step towards a more comprehensive analysis of human development index behavior may be provided by a univariate or a multivariate analysis, highlighting the tendency of countries  to develop similarities in structures, processes and performances over time. Multimodality can be easily detected by looking at Figure 1.

\subsection{Univariate analysis}

In order to better understand the dynamics of indicators measuring human development and to check for homogeneity within groups of countries, we first fit a univariate autoregressive model for the HDI. We aim at identifying  multiple behaviors. According to the model selection procedure described in Section 2.5, we identify three groups of countries interpreted as different process describing the evolution of HDI. The estimated distribution reveals the presence of “under-developed”, “middle-developed” and “developed” countries; their evolution is easily shown by looking at the evolution of the centroids of each group over time (Table 2). 

A clear and consistent pattern can be detected for groups' centroids, which show an interesting evolution over time. An increasing in the human development conditions is estimated; this can be considered as an indicator of groups' convergence toward better steady-states. We estimate an increment of the 11.29\%, 6.00\%, 4.41\% in the HDI group-mean values for each of the three groups respectively.

Focusing on HDI convergence for the identified clusters, we notice dynamic cluster-specific processes, i.e. clusters means move towards different steady-states. We are able to observe the implied transitions between clusters, occurring when memberships change. As expected (see Table 3), a few transitions occur and it is not possible to move from the “developed” to the “under-developed” cluster (and viceversa) from one year to another.

\subsection{Multivariate analysis}
Up to now, we have discussed about the dynamics (evolution over time and heterogeneity) in a univariate setting. Nevertheless, due to the complexity of the human development issue, a multivariate analysis should be pursued to provide further insights that univariate approaches cannot capture.
Furthermore, it is clear from Figure 2 that LEI, EI and GDPI are not independently distributed, necessitating a multivariate model specification.
Besides, when we face multivariate variables, the univariate approach is no longer sufficient and needs to be extended. In this context, we are likely to face complex phenomena which can be characterized by having a non-trivial correlation structure, and it is well known that, when responses are correlated, the univariate approach is less efficient than the multivariate one (see e.g. Davidson and MacKinnon, 1993).
The first evidence concerns the number of identified clusters (see Table 4): four well-defined clusters can be identified. Three of these are defined as before “under-developed”, “middle-developed” and “developed”; the fourth one shows a mixed behavior which is completely missing in the univariate analysis. In this cluster, the “mixed-developed” cluster, relatively small and time-decreasing values for the centroids are estimated for the LEI, with a simultaneous increasing of EI and GDPI centroids (see Table 5). Time-dependence can be easily detected by looking at the evolution of the centroids over time. A strong improvement in the estimated centroids values is obtained for the under-developed countries: LEI passes from 0.4838 to 0.5975; EI from 0.4559 to 0.5420 and GDPI from 0.4333 to 0.5014. Similarly, middle-developed and developed countries show an improved path for all the HDI components, even if such improvement is not dramatically relevant with respect to LEI and EI, indicating that GDPI represents the primary focus for these countries. Indeed, it is well-known that larger improvements from a low situation are more likely than improvements of the same size starting from a higher level.

According to the marginal membership matrix, we classify 10.41\% of countries in the “mixed-developed” cluster, while 15.93\%, 41.10\% and 32.56\% are clustered in the “under-developed”, “middle-developed” and “developed” clusters, respectively.
As described in Section 2, we assign countries to components according to the estimated matrix \textbf{U}. 
A graphical representation, which provides further insights on the obtained partitions, is provided in Figures 3-4.

Given this partition, we observe the implied transitions between components. Transitions are relatively rare during our sample period and a small number of countries account for most of them so that immobility rather than mobility is the norm. Of the 1144 possible transitions only 30 occur (2.62 \%; in line with the results of Paap and van Dijk, 1998, in a similar context). 117 countries in our sample remain assigned to the same component throughout the sample period. Of those that do transit from their initial component 22 shift just once, and 4, i.e. Cameroon, Congo, Honduras and Kenya,  shift more than once. Of the countries that never leave their initial component, 41 are those initially classified as developed countries; while 10 (Costa Rica, Croatia, Estonia, Honduras, Hungary, Latvia, Lithuania, Oman, Poland, Saudi Arabia, Slovakia) transit from the middle to the developed cluster. Bolivia and Nicaragua move from the mixed-developed to the middle-developed cluster; while Cambodia, Congo, Ghana, Myanmar and Papua New Guinea worsen their condition passing from the mixed-developed to the under-developed cluster. Kenya reaches the middle-developed cluster after visiting the mixed-developed and the under-developed ones; similarly Honduras, initially classified in the middle-developed cluster, moves in the mixed-developed cluster first and then, after returning in the middle-developed cluster, to the developed one. The empirical transition probability matrix is provided in Table 6.

\section{Concluding remarks}\label{s:disc}
In this paper, we presented a new clustering method, extending the well-known $K$-means algorithm, for the analysis of multivariate longitudinal data. The computational burden is dealt with a least squares approach through a coordinate descent algorithm.  Although many clustering strategies have been introduced using various perspectives, most of the
proposed approaches rely on very simple association structures. Here, we are able to distinguish and account for two different sources of association: {\it true} and {\it apparent}. In the former, actual and future centroids are directly
influenced by past values, which cause a substantial change over time in the
corresponding distribution. The latter case arises when statistical units are drawn from
heterogeneous populations. In the analysis of heterogeneous populations, ignoring time-related factors could produce a misleading statistical finding, since this unobserved but persistent heterogeneity can be interpreted as due to a
strong serial dependence. In some sense, our approach is semi-parametric. It shares the classical features of the K-means algorithm and, at the same time, a model-based structure is imposed on the evolution of the clustering over time.

The considered approach provides new insight when applied to the human development index. Human development is not the same and does not evolve in the same manner for all Countries. Furthermore, the rate of development is different, as well as different {\it steady state} of development can be reached by groups of Countries. The multivariate analysis provided in the empirical applications remarks the need of a more complex analysis than the one provided by simply looking at descriptive indexes.

It seems to be worthwhile to investigate the possibility of extending this approach to more complex problems. The inclusion of a dimensionality reduction step in the estimation step is straightforward. Least squares estimation allows us to include further independent covariates to estimate clusters centroids. Unbalanced panels can be easily treated, slightly modifying the proposed approach.

\newpage
\begin{table}\label{t:summ}\small
\begin{center}
\begin{tabular}{ccc|c|c|c|c|c|c|c|c}
&&\multicolumn{9}{c}{\textbf{Years}}\\\hline
& &\textbf{1997}& \textbf{1998}& \textbf{1999}& \textbf{2000}& \textbf{2001}& \textbf{2002}& \textbf{2003}& \textbf{2004}& \textbf{2005}\\
\hline
HDI & &&&&&&&&&\\
& Mean: & 0.715 & 0.719 & 0.723 & 0.729 & 0.735 & 0.739 & 0.744 & 0.749 & 0.761\\
& Std.Dev.: & 0.149& 0.149& 0.151&0.150&0.148&0.150&0.150&0.150&0.144\\
& Min: & 0.379 &0.382&0.395&0.403&0.422&0.425&0.450&0.445&0.452\\
& Max: &0.932 &0.935&0.939&0.942&0.944&0.956&0.963&0.965&0.968\\
\hline
LEI & &&&&&&&&&\\
& Mean: &0.710 & 0.713 & 0.706 & 0.708 & 0.711 & 0.712 & 0.717 &0.719 & 0.740\\
& Std.Dev.: & 0.155&0.156&0.167&0.169&0.175&0.179&0.180&0.182&0.158\\
& Min: & 0.240 & 0.260&0.250&0.250&0.170&0.150&0.120&0.100&0.265\\
& Max: & 0.920&0.920&0.930&0.930&0.940&0.940&0.950&0.950&0.954\\
\hline
EI & &&&&&&&&&\\
& Mean: & 0.783 & 0.788&0.793&0.798&0.807&0.812&0.816&0.817&0.823\\
& Std.Dev.: & 0.171&0.166&0.163&0.160&0.157&0.155&0.155&0.154&0.150\\
& Min: & 0.330&0.340&0.330&0.370&0.380&0.370&0.390&0.340&0.347\\
& Max: & 0.990&0.990&0.990&0.990&0.990&0.990&0.990&0.990&0.990\\
\hline
GDPI & &&&&&&&&&\\
& Mean: & 0.654&0.657&0.669&0.681&0.687&0.691&0.699&0.710&0.721\\
& Std.Dev.:&0.170&0.170&0.172&0.171&0.168&0.175&0.173&0.175&0.174\\
& Min:&0.310&0.310&0.320&0.250&0.340&0.330&0.350&0.360&0.370\\
& Max:& 0.960&0.970&1.000&1.000&1.000&1.000&1.000&1.000&1.000\\
\end{tabular}\end{center}\caption{Summary statistics.}
\end{table}

\begin{table}\label{t:centro}
\begin{center}
\begin{tabular}{c||c|c|c|}

&Under-developed&
Middle-developed&
Developed\\\hline
1997&
0,4861&
0,7128&
0,8739\\
1998&
0,4952&
0,7232&
0,8786\\
1999&
0,5004&
0,7306&
0,8841\\
2000&
0,5015&
0,7368&
0,8905\\
2001&
0,5048&
0,7407&
0,8890\\
2002&
0,5083&
0,7375&
0,8942\\
2003&
0,5099&
0,7375&
0,9002\\
2004&
0,5419&
0,7568&
0,9140\\
2005&
0,5410&
0,7556&
0,9125
\end{tabular}\end{center}\caption{Centroids evolution over time: HDI index.}
\end{table}

\begin{table}\label{t:transHDI}
\begin{center}
\begin{tabular}{c|c|c|c|}
&Under-developed&
Middle-developed&
Developed\\\hline
Under-developed&
0,9732&
0,0268&
0,0000\\
Middle-developed&
0,0157&
0,9686&
0,0157\\
Developed&
0,0000&
0,0134&
0,9866
\end{tabular}\end{center}\caption{Transition matrix: HDI index. (time t-1 on the rows; time t on the columns)}
\end{table}

\begin{table}\label{t:modsel}
\begin{center}
\begin{tabular}{c|c}
Number of clusters & CH\\
\hline
2 & 1670.679\\
3 & 1713.421\\
4&1769.209\\
5&1540.763\\
6&1479.506
\end{tabular}\end{center}\caption{Model selection}
\end{table}

\begin{table}\label{t:centr}
\begin{center}
\begin{tabular}{c|c|c|c|c|}
\multicolumn{5}{c}{LEI}\\
\hline
&Under-developed&
Middle-developed&
Developed&
Mixed-developed\\
1997&
0,4838&
0,7485&
0,8588&
0,5255\\
1998&
0,4932&
0,7522&
0,8572&
0,5205\\
1999&
0,4900&
0,7418&
0,8573&
0,4042\\
2000&
0,4864&
0,7430&
0,8580&
0,3882\\
2001&
0,5170&
0,7498&
0,8613&
0,3354\\
2002&
0,5209&
0,7527&
0,8626&
0,3169\\
2003&
0,5470&
0,7455&
0,8714&
0,3062\\
2004&
0,5939&
0,7665&
0,8823&
0,4172\\
2005&
0,5975&
0,7579&
0,8780&
0,3916\\ \hline\hline
\multicolumn{5}{c}{EI}\\
\hline
&Under-developed&
Middle-developed&
Developed&
Mixed-developed\\
1997&
0,4519&
0,8318&
0,9168&
0,7050\\
1998&
0,4727&
0,8334&
0,9198&
0,7168\\
1999&
0,5012&
0,8243&
0,9218&
0,7533\\
2000&
0,5092&
0,8254&
0,9230&
0,7755\\
2001&
0,5196&
0,8372&
0,9280&
0,7462\\
2002&
0,5296&
0,8407&
0,9340&
0,7369\\
2003&
0,5383&
0,8416&
0,9404&
0,7246\\
2004&
0,5486&
0,8692&
0,9518&
0,7497\\
2005&
0,5420&
0,8532&
0,9359&
0,7284\\
\hline\hline
\multicolumn{5}{c}{GDPI}\\
\hline
&Under-developed&
Middle-developed&
Developed&
Mixed-developed\\
1997&
0,4333&
0,6198&
0,8729&
0,5400\\
1998&
0,4368&
0,6273&
0,8667&
0,5295\\
1999&
0,4456&
0,6328&
0,8718&
0,5575\\
2000&
0,4600&
0,6375&
0,8787&
0,5918\\
2001&
0,4743&
0,6441&
0,8798&
0,5792\\
2002&
0,4704&
0,6438&
0,8883&
0,5900\\
2003&
0,4791&
0,6510&
0,8890&
0,5885\\
2004&
0,5055&
0,6724&
0,9176&
0,5795\\
2005&
0,5014&
0,6694&
0,9119&
0,5799\\\hline
\end{tabular}\end{center}\caption{Centroids evolution over time: LEI, EI and GDPI.}
\end{table}

\begin{table}\label{t:transMV}
\begin{center}
\begin{tabular}{c|c|c|c|c|}
&Under-developed&
Middle-developed&
Developed&
Mixed-developed\\\hline
Under-developed&
0,9672&
0,0000&
0,0000&
0,0328\\
Middle-developed&
0,0000&
0,9767&
0,0211&
0,0021\\
Developed&
0,0000&
0,0000&
1,0000&
0,0000
\\
Mixed-developed&
0,0583&
0,0500&
0,0000&
0,8917\\\hline
\end{tabular}\end{center}\caption{Transition matrix: multivariate analysis. (time t-1 on the rows; time t on the columns)}
\end{table}

\newpage
\begin{figure}[h]
\begin{center}\includegraphics[scale=.4]{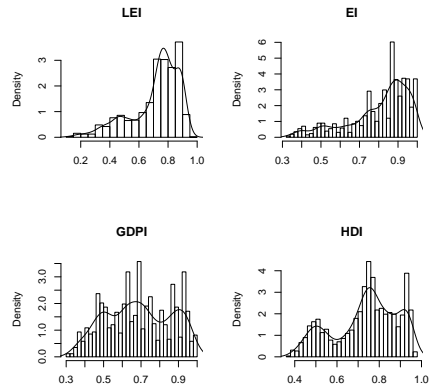}\label{f:datadescr}\caption{LEI, EI, GDPI and HDI marginal distributions.}\end{center}
\end{figure}

\begin{figure}[h]
\begin{center}
\includegraphics[scale=.45]{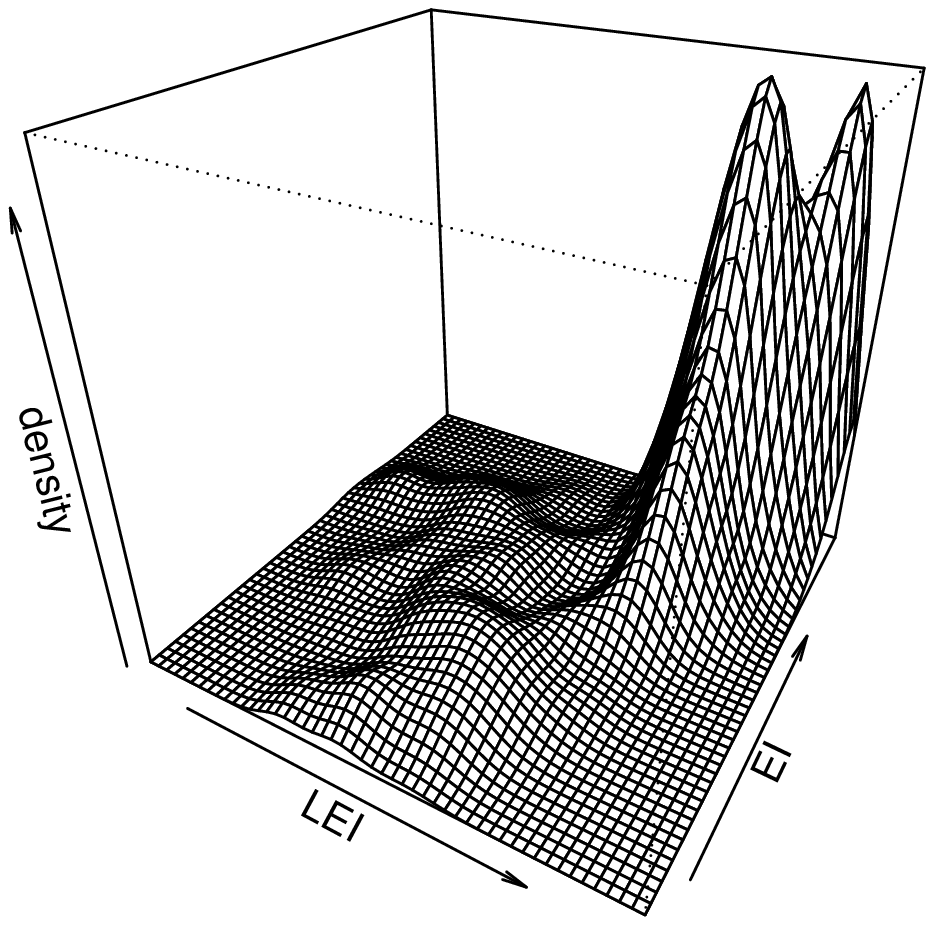}
\includegraphics[scale=.45]{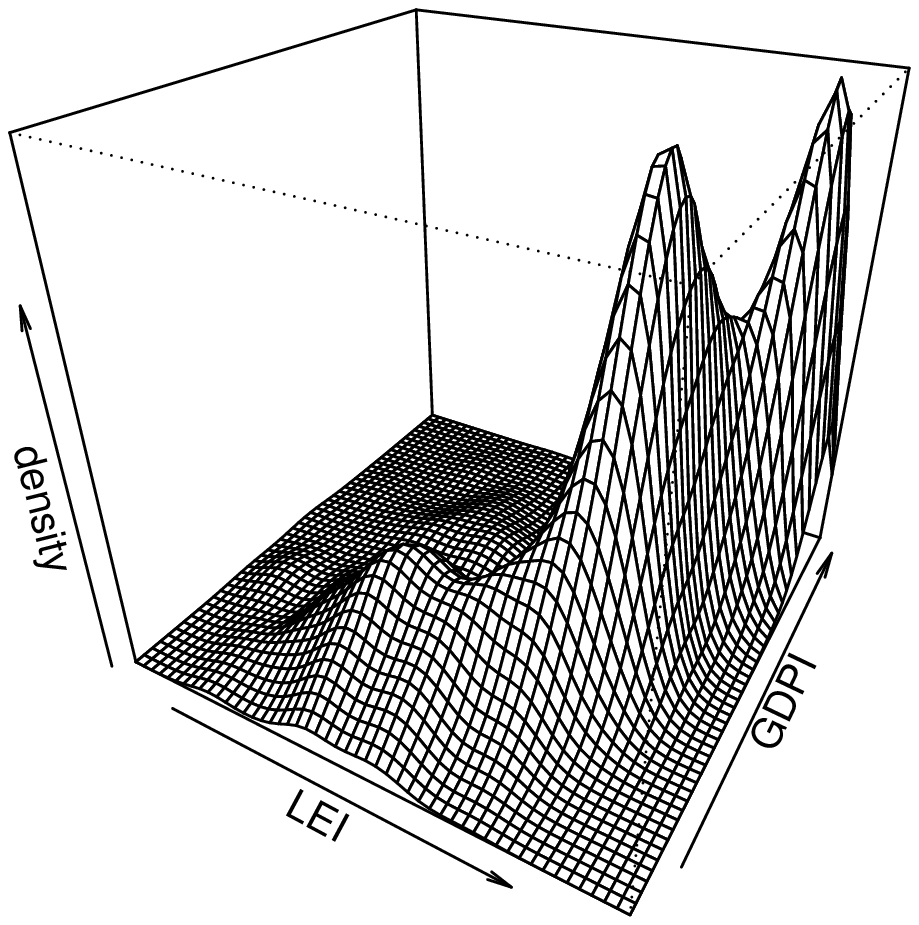}\\
\includegraphics[scale=.45]{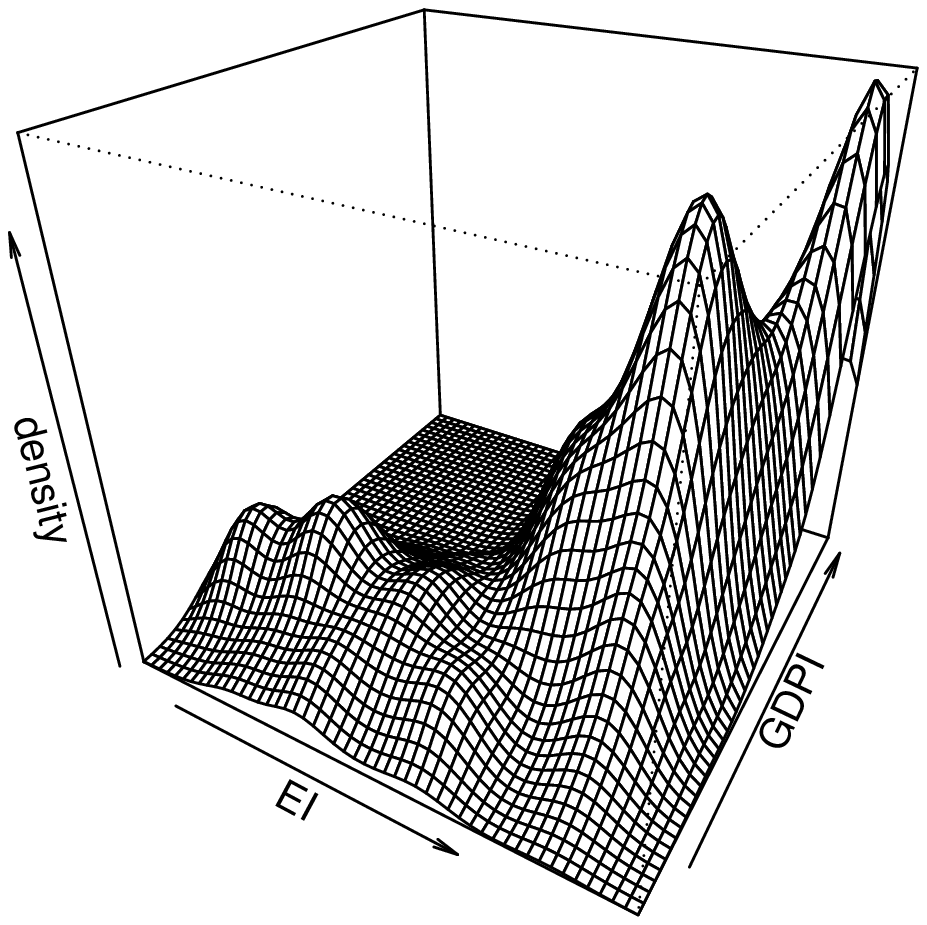}\label{plot}\caption{LEI, EI and GDPI relationships.}\end{center}
\end{figure}

\begin{figure}[h]\small
\centering{
\subfigure[LEI vs. EI]{\includegraphics[scale=.45]{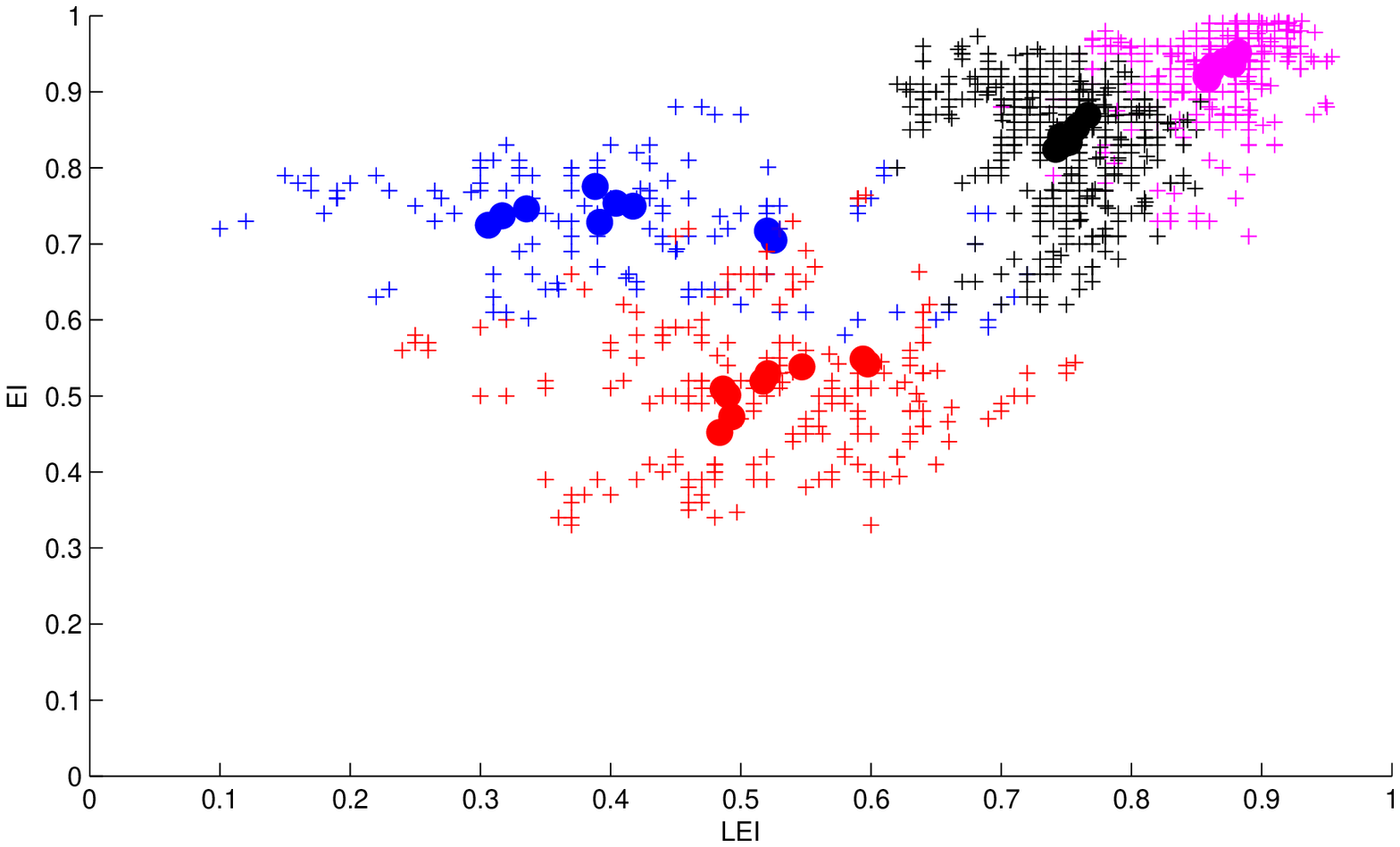}}
\subfigure[LEI vs. GDPI]{\includegraphics[scale=.45]{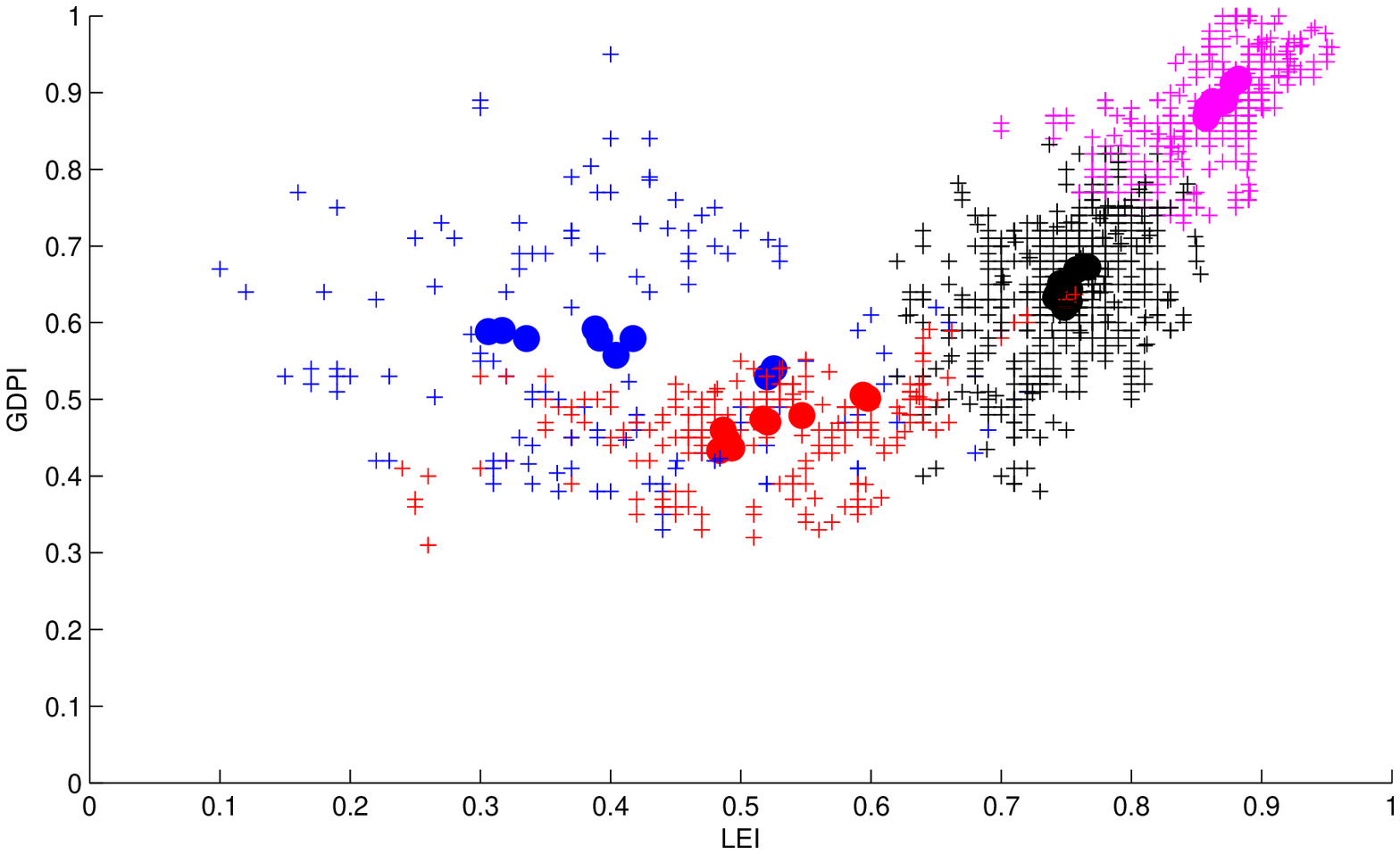}}
\subfigure[EI vs. GDPI]{\includegraphics[scale=.45]{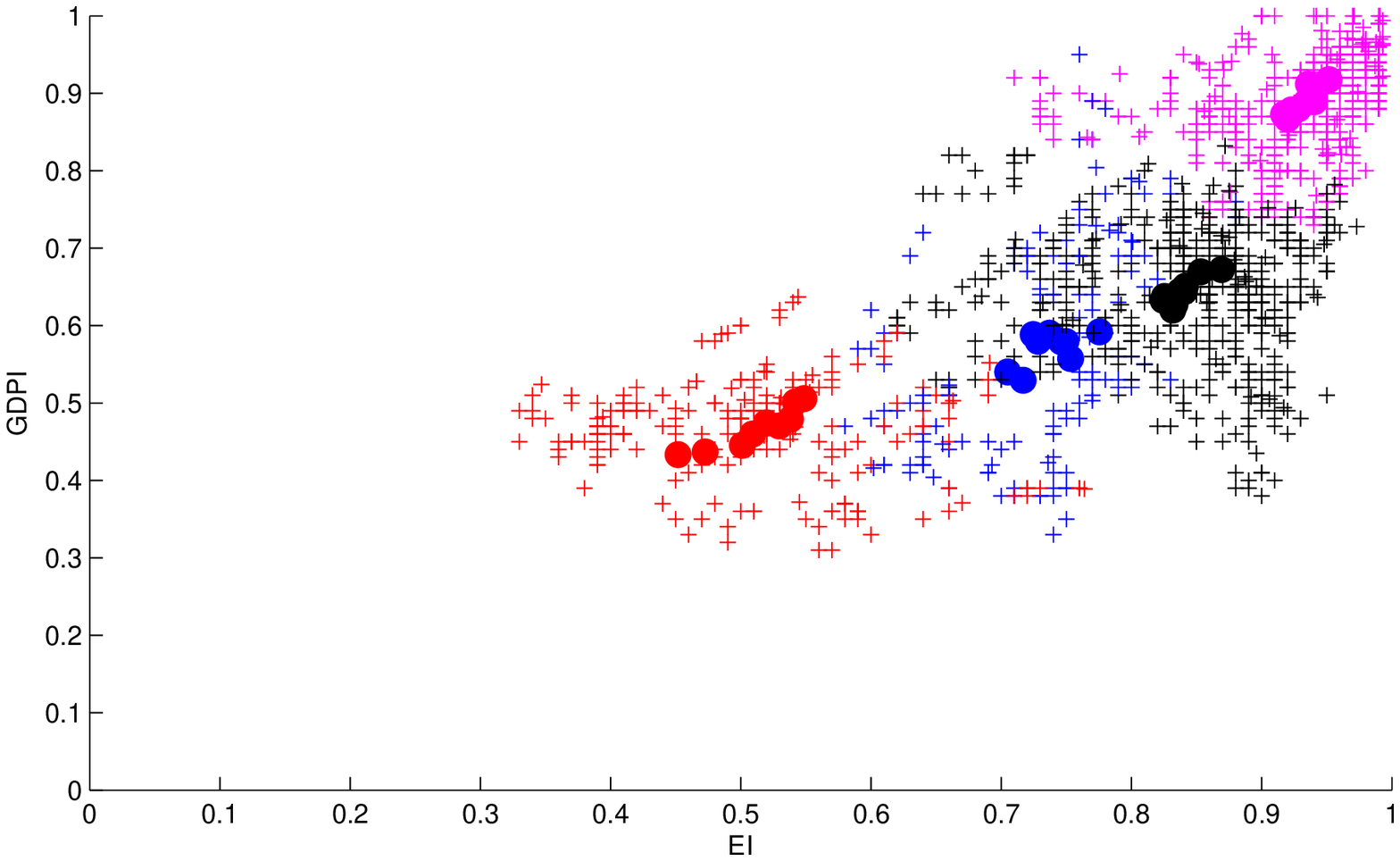}}
}\label{bid}\caption{Bidimensional clustering: LEI, EI and GDPI relationships. Dots represents groups centroids. Blue = Mixed-development; Red = Under-developed; Magenta = Developed; Black = Middle-developed}
\end{figure}

\begin{figure}[h]\small
\centering{
\includegraphics[scale=.65]{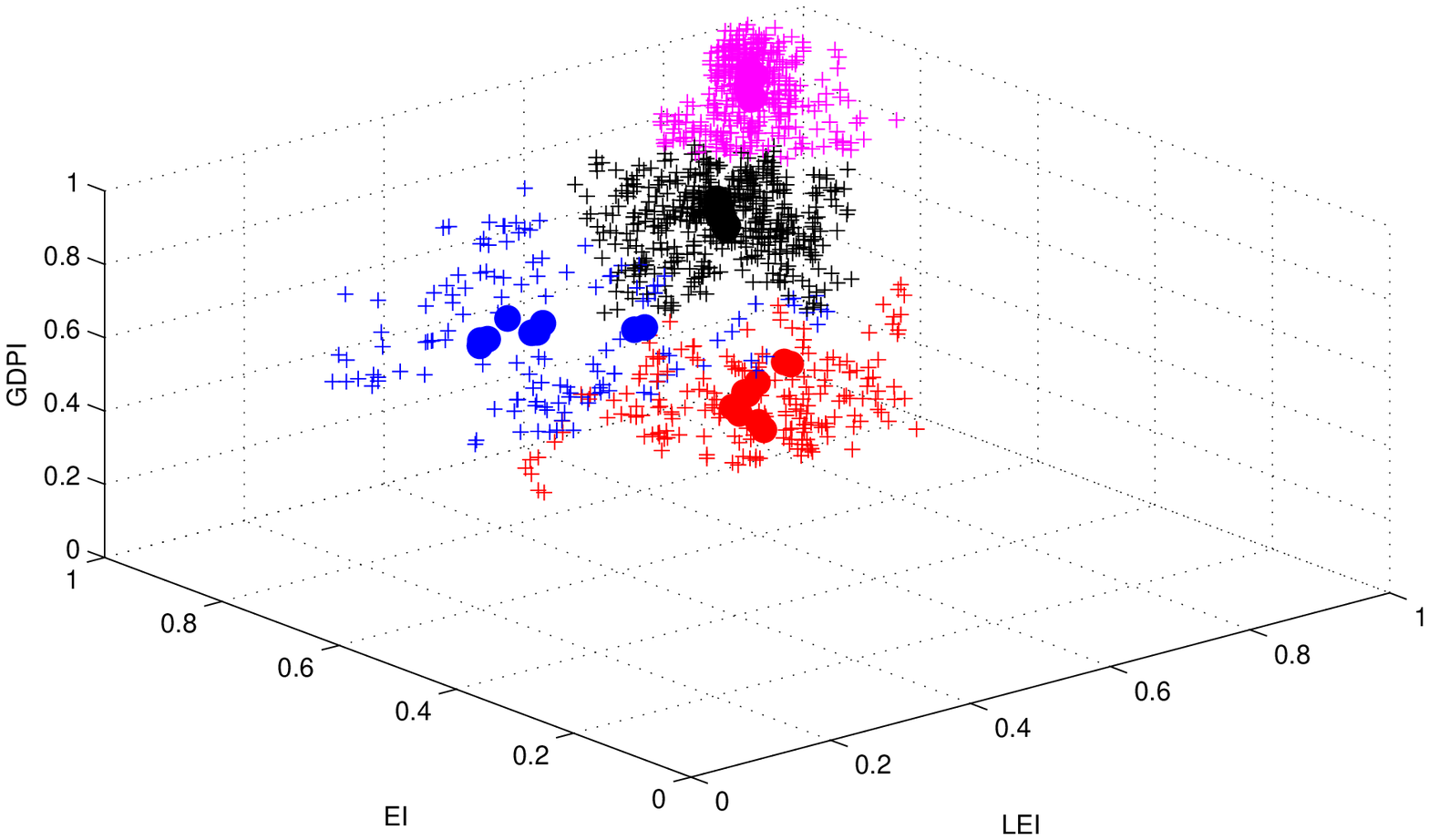}
}\label{trid}\caption{Tridimensional clustering: LEI, EI and GDPI relationships. Dots represents groups centroids. Blue = Mixed-development; Red = Under-developed; Magenta = Developed; Black = Middle-developed}
\end{figure}
\end{document}